# Non-equilibrium dynamics of the glass transition: new perspectives from colloidal hard spheres

W. van Megen[1] and H. J. Schöpe[2]

[1]Department of Applied Physics, Royal Melbourne Institute of Technology, Melbourne, Victoria 3000, Australia

[2]Institut für Angewandte Physik, Universität Tübingen, Auf der Morgenstelle 10, 72076 Tübingen, Germany

ABSTRACT

The occurrence of caging distinguishes a super-packed fluid from one in thermodynamic equilibrium and is pivotal for the process of vitrification; the formation of the amorphous solid that occurs on super-packing. Determination of the spatial extent of caging and the fraction, $F_{cage}$, of caged particles, from several light scattering experiments on suspension of hard spheres, form the basis of a proposed framework that expresses dynamical heterogeneity in terms of a dynamical solid identified by collective dynamics of caged particles. The dynamical solid is observed in a spatial window of width, $\Delta q$, centred about the position, $q_m$, of the structure factor maximum, that is separated from the dynamical fluid of purely Brownian currents in the complementary window. Equality of $F_{cage}$ and the reciprocal space, $\Delta q/q_m$, where caging dynamics manifests is a pivotal result of this study. It places a lower limit, $q_m-\Delta q$, on the spatial frequency of correlated cage fluctuations and sets their time scale: In effect $F_{cage}$, determines the mechanism(s) by which the time correlation functions in the super-packed hard sphere suspension decay. Central to this decay are non-stationary, intermittent correlated cage fluctuations of the β-regime in which we identify three dynamical scenarios: (i) non-propagating cage fluctuations, for $F_{cage}<1/2$, (ii) propagating and non-propagating cage fluctuations respectively to and from a plateau, for $½<F_{cage}<1$, (iii) propagating cage fluctuations to a plateau, for $F_{cage}=1$. Once the fraction of caged particles exceeds ½, in regime (ii) the dynamics are redolent of the phononic and non-phononic modes exposed in density of states analyses and the critical decay and the von Schweidler law of the mode-coupling theory. These mechanisms characterise the reversible processes in the super-packed suspension. From their perspective, we gain insight into irreversible, aging processes and the first steps in the separation of the crystalline phase.

# I. INTRODUCTION

When a liquid is cooled below its freezing point quickly enough to bypass crystallisation its structural relaxation time and resistance to flow increase sharply and can do so to such an extent that, with sufficient super-cooling, the liquid vitrifies. The physics of this phenomenon continues to be one of the more intriguing and studied aspects of classical condensed matter[1-11]. Where the rigidity of a crystalline solid, on account of its lattice modes, is a direct consequence of its structure, so too one might expect the rigidity of the amorphous solid to derive from its structure. It's puzzling, therefore, that vitrification occurs with merely subtle changes in structure.

The long-standing picture of this process is, on super-cooling, or super-packing, crowding causes particles to be caged by their neighbours. There is an attendant separation of solid-like regions of particles seemingly confined to their neighbour cages and fluid-like regions of particles that roam relatively freely[12-18]. This dynamical heterogeneity can be expressed by a length, $L_D$, of the average size of assemblies of caged particles that can only rearrange in some cooperative fashion – an idea first proposed by Adam and Gibbs[19]. In a variant of this concept, introduced by Mountain[20, 21], $L_D$ represents the maximum wavelength for which the super-cooled fluid is viscoelastic. In either case, increasing super-cooling supposedly increases the fraction of caged particles and $L_D$ to a point where all particles are caged resulting, at least in an ideal sense, in a macroscopically rigid amorphous assembly of particles rattling in their respective neighbour cages[18, 22-26]. However, Mountain's definition connects $L_D$ with phenomenology in that it presents a dynamical boundary between viscous flow for wavelengths $\lambda > L_D$ from elastic restoration for wavelengths $\lambda < L_D$; put simply, $L_D$ separates the dynamics characteristic of an elastic solid from that of a viscous fluid. One question this raises is, to what extent is delayed structural relaxation of the super-cooled fluid due to the elastic memory of the "solid" and due to any increase in viscosity of the "fluid"?

The preceding suggests a connection between caging, dynamical heterogeneity and, by the presence of long wavelength ($\approx L_D$) transverse currents, sluggish structural relaxation expressed by some time scale $\tau_0$. However, a quantitative connection between $\tau_0$ and $L_D$ and, for that matter, the extent of caging, as might be expressed by the fraction, $F_{cage}$, of particles that are caged, has not yet been established. Moreover, given these features pertain to super-cooled fluids, we may also ask if a fluid in thermodynamic equilibrium, by its inability to support a shear stress, is dynamically homogeneous and devoid of caging?

Microscopic studies expose rather more directly not just the isotropic rattling of caged particles but also, by way of correlated reversals in the direction of motion[23, 27], the initial stages of cooperation that may extend to many particles[28-31]. Not implausibly, these observations expose aspects of more complex and extensive collective processes; non-phononic and phononic[32-36] modes, identified respectively with structural relaxation and the Boson peak in deeply super-cooled fluids and glasses, not directly exposed under microscopic observation. While these modes have neither the density of states nor the well-defined spatial or temporal frequencies that characterise the lattice modes of a crystalline solid, they are effective, nonetheless, in supporting an applied stress[13, 21, 34] and, by the same token, impair flow. It also appears that structural relaxation proceeds by intermittent jumps of either single particles[18, 37-43] or involving the cooperation of many particles[22, 25]. Whether or how these are connected to non-phononic and/or phononic processes and, moreover, which are reversible or irreversible, is not clear.

Structural relaxation is read more traditionally from the decay of the time correlation function of the particle number density, or intermediate scattering function (ISF)[2, 3, 44-46] which expresses the de-phasing of the scattered light, or density, field. The multiple stages of this decay seen generally for super-cooled fluids can be expressed by the following sequence of mechanisms; (I) intra-cage rattling, or collisions between caged particles which set in train collective motions comprising, (II) the processes in the $\beta$ regime, followed by (III), the $\alpha$ process. We anticipate that delay in structural relaxation, usually identified with the $\alpha$ process, is not due to infrequent "escapes" of particles from their respective cages per se, but rather by delay in the dephasing of the amorphous structure through correlated cage fluctuations; ie, anisotropic collective dynamics (II) within the structure[47, 48]. This being the case, then how are these collective processes, be they phononic or non-phononic, manifested in the $\beta$ regime? And reiterating the question above, in what way do these relaxation mechanisms, ostensibly unique to the super-cooled fluid, differ from those of a thermodynamically equilibrated fluid?

In addressing the above issues and questions in this paper we are guided by the above notion that the distinguishing characteristics of super-cooled/packed fluids stem from the presence of regions of caged particles and the rattling motion that that caging incurs. Accordingly, we seek to establish a connection between these characteristics – the increasing/diverging length and time scales, the collective manifestations of caging expressed in (II) and (III) above and the scaling laws they satisfy – and the manner of their connection to a common, quantifiable property or order parameter.

Our enquiry is based on re-examination of results of dynamic light scattering (DLS) experiments on suspensions of particles with hard-sphere like interactions[49-53]. Complementary results from molecular dynamics (MD)[54, 55] and confocal microscopy[23, 56] are also relied upon. The system of hard spheres is an ideal but nonetheless valuable reference system that has over many decades enhanced our understanding of condensed matter considerably[57, 58]. Its only control parameter is the packing fraction, $\phi$, in terms of which fluid and crystal phases coexist at $\phi_f$=0.494 and $\phi_m$=0.545[59, 60]. As well as this freezing/melting transition a hard sphere suspension also displays a glass transition (GT) which, in view of its accessibility by optical techniques, has been studied extensively[9, 44]. However, its location, $0.56 \lesssim \phi_g \lesssim 0.6$, on account of irreversible, aging processes, is not only imprecise but also contentious[61-66]. We circumvent both issues to an extent by isolating, in the first instance, just the reversible processes and thereby exposing a transition, at approximately $\phi_g \approx 0.570$, to an underlying *ideal* (ageless) glass. The transition in this so-called "fragile" glass former[67] is effected by excluded volume effects and the various phenomena, mentioned above, associated with the transition stem from the caging of particles rather than the intricacy of a network of bonds that underpins the phenomenology of more familiar "strong" glass formers.

Concerning the question of how a suspension in thermodynamic equilibrium can be distinguished from one that is metastable, a preceding paper[68] considered a hard-sphere suspension from the perspective of the suspending liquid (solvent). This for the reason suspended hard sphere particles interact only hydrodynamically – by solvent mediated longitudinal momentum currents which transmit, at sonic speed, energy exchanges between the particles[69-71]. Having lost memory of those exchanges, expected on the time scales of standard DLS experiments, particles diffuse independently of one another. On these time scales dynamical variables, such as the longitudinal particle current density, manifest as randomly fluctuating, Gaussian variables, and their corresponding time correlation functions comprise superpositions of (independent) exponential decays. Thus, whatever stretching the particle current auto-correlation function (CAF), for example, exhibits necessarily results from a spread in the decay times of those exponential decays rather than from some intrinsic collective dynamics. This is the dynamics – Brownian motion – that obtains in the thermodynamically equilibrated, single phase suspension ($\phi < \phi_f$). Accordingly, deviations in the decay of the CAF from a stretched exponential, seen in the super-packed ($\phi > \phi_f$) suspension[48], indicate the presence of interactions among particles consequent on caging.

Our study is based in large part on the CAF for it, rather than the ISF, more usually considered in studies of the dynamics of suspensions[44, 72], gives the more direct exposure of the collective consequences of caging. The reason is that contributions to the ISF that follow exponential or weakly stretched exponential functions of the delay time tend to be weakly exposed, and may even be suppressed to the noise floor, in the CAF. At the same time, those processes that are expressed by power-laws tend to be more prominently exposed. The evidence to date indicates that the α decay and aging – respectively reversible and irreversible relaxation processes – can be approximated by weakly stretched exponential terms in the ISF[73-75], and the processes in the β regime, describing collective caging dynamics, by power laws[75]. So, in a nutshell, α relaxation and aging tend to be more prominently exposed by the ISF, whereas the CAF is more sensitive to the collective expressions of caging. Accordingly, we quantify the decay of time correlation functions in terms of stretched exponentials and power laws, then provide interpretation in terms of the above three mechanisms (I-III) and the nature of qualitative changes in those mechanisms, firstly on traversal of the thermodynamic freezing point (at $\phi_f$), then, on super-packing, approach to and traversal of the (ideal) GT.

This paper is organised as follows: Definitions of the relevant dynamical properties and brief descriptions of the hard sphere suspensions and light scattering procedures are outlined in the following Section. Results and Discussion in Sec. III. comprise the bulk of the paper, opening in Sec. III.A with the signature of caging, as read from the CAF, and how that signature distinguishes the metastable/super-packed suspension from one in thermodynamic equilibrium. In Sec. III.B we present a dynamical framework for vitrification. This framework defines dynamical heterogeneity by the separation of longitudinal particle currents that carry the signature of caging, in the window (~2Δq) of spatial frequencies centred on the position, $q_m$, of the main peak of the structure factor, from Brownian currents in the complementary window. In Sec III.C we estimate $F_{cage}$; first, from DLS (Sec. III.C.1); second, from molecular dynamics (Sec. III.C.2); third, from confocal microscopy (Sec. III.C.3). Examination of the dynamics of the "solid" of caged particles comprises discussion of the local rattling in Sec. III.D.1, followed in Sec. III.C.2 by the collective manifestations of rattling identified by the power law decays that define the β-regime. Sec. III.D tests the results of this framework against measured decay times of the α-process. In Sec. III.F we aim to place the reversible cage dynamics of the amorphous solid in the broader context with the irreversible processes. Sec. IV briefly discusses a few issues that, in our view, warrant further investigation. Summary and conclusions follow in Sec. V.

## II. METHODS

The usual quantity obtained by DLS is the intermediate scattering function (ISF)[72, 76]

$$f(q,\tau) = \langle \rho(q,0)\rho^\dagger(q,\tau)\rangle / \langle|\rho(q)|^2\rangle \qquad (1)$$

where

$$\rho(q,t) = \sum_{k=1}^{N} \exp[-i\mathbf{q}\cdot\mathbf{r}_k(t)] \qquad (2)$$

is the $q^{th}$ spatial Fourier component of the particle number density, $\tau$ is the delay time, $\mathbf{r}_k(t)$ the position of particle k at time t and "†" denotes the complex conjugate. Brownian particle number density fluctuations are characterised by the short-time diffusion coefficient $D(q)=d/d\tau[\ln(f(q,\tau\rightarrow"0")/q^2])$, where "0" refers to the lower detection limit. The time auto-correlation function of the longitudinal particle current[57] is defined by

$$C(q,\tau) = q^2 \langle j(q,0)j^\dagger(q,\tau)\rangle / \langle|\rho(q)|^2\rangle = -d^2 f(q,\tau)/d\tau^2, \qquad (3)$$

where

$$j(q,t) = \sum_{k=1}^{N} \hat{\mathbf{q}}\cdot\mathbf{v}_k(t)\exp[-i\mathbf{q}\cdot\mathbf{r}_k(t)] \qquad (4)$$

and $\mathbf{v}_k(t)$ is the velocity of particle k at time t. A complementary property is the self ISF,

$$f_s(q,\tau) = \langle \exp[i\mathbf{q}\cdot\Delta\mathbf{r}(\tau)]\rangle, \qquad (5)$$

where $\Delta \mathbf{r}(\tau)$ is the displacement of a "tagged" particle in the time interval $\tau$. The mean-squared displacement (MSD) is

$$\langle \Delta r(\tau)^2 \rangle = \lim_{q \to 0} -\frac{6}{q^2} \ln(f_s(q,\tau)), \tag{6}$$

from which we obtain the stretching index, $\mu$, defined here by the minimum logarithmic slope of the MSD,

$$\mu = \min[d \log(\langle \Delta r^2(\tau) \rangle)/d \log(\tau)]. \tag{7}$$

Our interpretation of the dynamics relies on whether the decay of a time correlation function can be expressed by a stretched exponential (SE) function,

$$\psi(\tau) = A \exp[-(\tau/\tau_{se})^\gamma], \tag{8}$$

or by a power law

$$\psi(\tau) = B\tau^\nu, \tag{9}$$

of the delay time. The SE comprises the superposition[77, 78]

$$\psi(\tau) = \int g(\tau_t) \exp[-\tau/\tau_t] d\tau_t \tag{10}$$

of exponential functions weighted by the distribution, $g(\tau_t)$, of decay times, $\tau_t$. When this holds for the ISF then, since the derivatives (Eq. (3)) must be taken inside the integral, it also holds for the CAF.

The results derive from DLS measurements[48-51, 74, 79] of several suspensions, comprised of either polymer particles with thin, oligeromised surface coatings (suspension P) or cross-linked microgel particles (suspensions M1, M2) (Table I), suspended in solvents that maintain integrity of the particles and provide the refractive index matching that minimise multiple scattering. Significantly, particle interactions are short-ranged and steeply repulsive; hard sphere like. Preparation of suspensions that yield the self ISF (Eq. (5)) has been achieved by adding to the suspension of polymer particles a small concentration of identically sized and coated silica particles (suspension T), and then adjusting the refractive index of the mixture so that scattering from its structure is suppressed [80, 81]. In all cases the packing fraction, $\phi$, has been determined by referencing the observed equilibrium colloidal fluid-crystal separation to that of the ideal hard sphere system[60, 82]; setting the freezing value at that, $\phi_f=0.494$, of the perfect hard sphere fluid absorbs dependence of the actual freezing value on suspension specific characteristics such as polydispersity and surface properties of the particles. In addition to the first order freeing/melting transition, a glass transition has been located at $\phi_g \approx 0.57$[46, 75, 83]. Properties of the particles, in addition to those listed in table I, and light scattering procedures are documented elsewhere[49, 52, 53, 82, 84]. In the results presented below the spatial frequency, q, and all lengths are expressed in terms of the particle radius, R, and delay times in units of the Brownian time, $\tau_B$ (Table I).

The CAF has been obtained here by numerical differentiation of the ISF (Eq. (3)). Using Savitzky Golay smoothing, each data point of the ISF is assigned a polynomial fitted symmetrically to the data around this point. By analytical differentiation of the polynomial the value of the derivative for the data point is calculated.

|  | Suspension P[50] | Suspension T[49] | Suspension M1[51] | Suspension M2[53] |
|---|---|---|---|---|
| Radius, R | 185nm | 200nm | 430nm | 370nm |
| Polydispersity, $\sigma$ | 8% | 6% | 4% | 2% |
| $\phi_m$ | 0.530±0.002 | 0.536±0.003 | 0.540±0.002 | 0.545±0.002 |
| Brownian time, $\tau_B$ | 0.013s. | 0.016s. | 0.175s. | 0.111s |

TABLE I. **Suspension properties.** Suspensions P, polymer particles; T, mixture of polymer and silica particles; M1, M2; microgels. $\tau_B = R^2/(6D_0)$, $D_0$ is the diffusion coefficient of a freely diffusing particle.

There have been other DLS studies of the GT dynamics of hard sphere like colloidal systems[65, 85-87]. However, these have not determined the CAFs nor have they noted the first order freezing transition which, in the context of the GT, may not have been considered relevant. We consider otherwise because knowing the location of this transition allows dynamical signatures, traceable to caging, to be unambiguously associated with the metastable, super-packed state. At the same time, on account of the suspensions' polydispersity (Table I), formation of colloidal crystals is slow enough for the dynamics of the metastable states to be studied[88, 89]. In the case of suspension M2 the induction time for crystallisation is shortest due to the narrow particle size distribution. This limits determination of time correlation functions in the metastable state to $\phi \lesssim 0.53$. The extent polydispersity affects the miscibility gap, $\phi_m - \phi_f$, of our suspensions can be inferred from estimates of $\phi_m$ (Table I).

As mentioned in Sec. I, the CAF allows us to identify an idealised GT for hard spheres. Diverging relaxation times are implicit. In reality, of course, such divergences are rounded by differential arrest stemming from broad and/or asymmetrical particle size distributions and aging[61, 63, 64, 85]. Fortuitous rather judicious "choice" of nearly symmetric particle size distributions[90] in the present case delays separation of the crystal phase[89, 91], limits differential arrest and, along with suppression of aging in the CAF [92], allows identification and definition of an idealized GT. Thus, the decay of the CAF, being dominated by reversible processes, is to good approximation invariant to time translation. We note that the latter property of time correlation functions is a necessary indicator of thermodynamic equilibrium. However, it is not sufficient. As we find below, it's the *manner* in which these correlation functions decay, rather than their decay to zero as such, that tells us whether or not the suspension *is* in thermodynamic equilibrium.

## III. RESULTS AND DISCUSSION

### A. Distinction between equilibrium and non-equilibrium.

First, we show that the dynamics of a metastable suspension differ from one in thermodynamic equilibrium and that the CAF is more sensitive than the ISF in exposing that difference. Fig. 1 shows ISFs and CAFs for packing fractions, $\phi=0.475$ and $\phi=0.519$, approximately 5% either side of the freezing point, $\phi_f=0.494$, at a spatial frequency corresponding to the position, $q_m \approx 3.5$, of the main peak of the static structure factor. ISFs (Fig. 1a) appear as smooth concave functions of $\tau$ that can be approximated by SEs (Eq. (8)); aside from differences in their decay rates there is no qualitative difference between them. For the lower $\phi$ ($<\phi_f$) the corresponding CAF (Fig. 1b) can also be approximated by a SE. But for the larger $\phi$ ($>\phi_f$) the CAF displays variations in curvature not evident in the ISF. On increasing the packing fraction to $\phi=0.54$ (Fig. 2) one sees that even the ISF cannot be fairly approximated by a single SE, but the decay from a height, $A \approx 0.8$, can and does so in a manner characteristic of the α process[2, 3, 45]. Note however, that the decay preceding this, $f(q_m, \tau \lesssim \tau_{3x})$, is more difficult to quantify; neither plateau nor power law is evident. At the same time, in the corresponding CAF the α process, now far less prominent, is clearly preceded by a decay that can be approximated by a power law in $\tau$. The various time scales used in this paper, such as $\tau_{3x}$ introduced here, are compiled in Table II at the end of this paper.

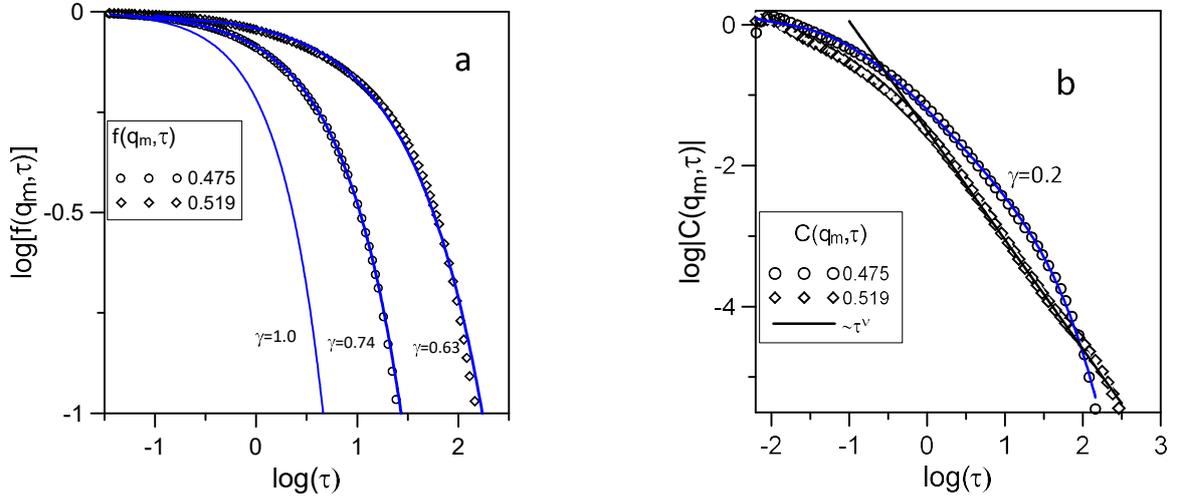

FIG. 1. ISFs in panel (a) and corresponding CAFs in panel (b) for packing fractions indicated. The results are for suspension P and spatial frequency $q=q_m \approx 3.5$. The solid (blue) lines are SEs (Eq. (8)), with stretching indices, $\gamma$, indicated. The single exponential ($\gamma=1$) in (a) is included for reference. In (b) the black line is a power law, $\sim \tau^\nu$, with $\nu=-1.5$. CAFs are scaled vertically to coincide at left limit of the ordinate.

We see by these variations in curvature that the CAFs (Fig. 1b and 2) of the metastable suspension expose processes not apparent in the ISFs. We attribute these curvature variations to caging – perceiving this in a dynamical sense as impairment of the putative Brownian motion of some fraction, $F_{cage}$, of particles due to close proximity of their neighbours. Reading of these non-Brownian, deterministic dynamics from the data is the main focus of what follows.

To spell out further the difference just illustrated between the thermodynamically stable and metastable suspensions we recall (Sec. I), that for packing fractions less than the freezing value, $\phi_f$, the CAF can be expressed by a SE (Eq. (8)). Stretching in these cases results from a superposition of exponential decays (Eq. (10)), each expressing the time correlation function of an *independent*, random Gaussian (Brownian) current . (Strictly speaking, the integral in Eq. (10) represents a sum over these independent exponentials.) Moreover, the stretching index, $\gamma$, shows no systematic dependence on q over the experimental range, $1 \lesssim q \lesssim 5$, of spatial frequencies bracketing $q_m(\approx 3.5)$ [50, 51, 68]. This tells us that the spread of diffusivities is independent of the spatial frequency. The crucial point is that the SE decay of the CAF is a necessary, albeit but not sufficient, signature of Brownian motion; the dynamical signature of a suspension in thermodynamic equilibrium.

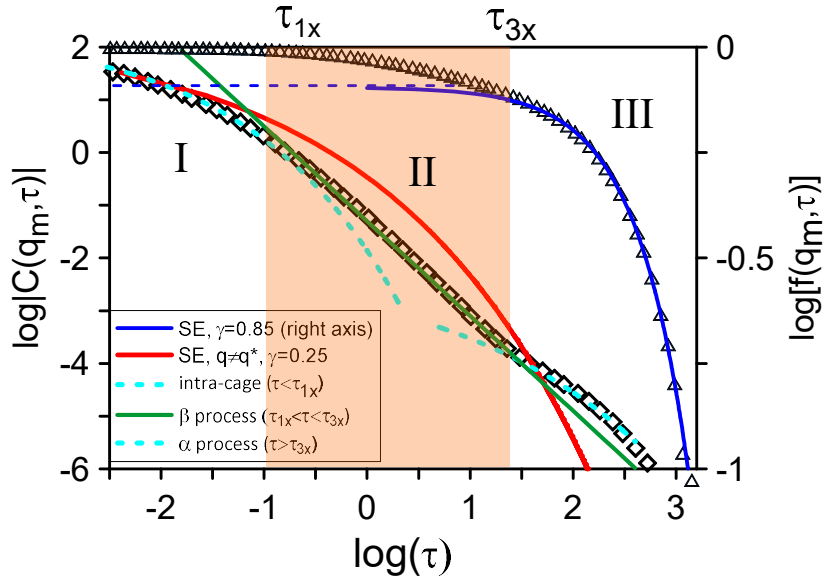

FIG. 2. DLS results (Suspension P); ISF (triangles) and corresponding CAF (squares) for $\phi=0.540$ and $q=q_m$. The ISF and the SE (blue line, Eq. (8) for A=0.8, $\gamma=0.85$) refer to the right axis. In the CAF we identify the following sequence of relaxation mechanisms; (I) intra-cage rattling for $\tau \lesssim \tau_{1x}$; (II) the shaded area is the β-regime where, for this case, the decay is approximated by the power-law, $\sim \tau^\nu$, ($\nu \approx -1.8$, green line) for $\tau_{1x} \lesssim \tau \lesssim \tau_{3x}$; (III) the α-process for $\tau \gtrsim \tau_{3x}$. The red line is the SE ($\gamma \approx 0.25$) that applies for all CAFs for $\phi_f < \phi \lesssim \phi_g$ and $q \neq q^*$ as defined in the text. The dashed curves are not optimised SE fits to the data but serve to highlight the difference between the power-law approximation to the decay by mechanism (II) in the β-regime from the curvature of the decays by mechanisms (I) and (III) outside the β-regime.

Reiterating, deviations from a SE that cannot be absorbed by experimental noise, evident in the super-packed region ($\phi > \phi_f$) (Fig. 1b and 2), indicate that there are non-Brownian, deterministic

processes – interactions/collisions – on experimental time scales ($\tau>10^{-6}$s). Various DLS experiments[48, 50, 51, 79] find this signature of caging emerges as $\phi$ just exceeds $\phi_f$ for spatial frequencies, $q^*$, close to $q_m$. On further super-packing, this signature spreads more or less symmetrically over a widening spatial window, $q_m \pm \Delta q$, defining $q^* \in \{q_m-\Delta q, q_m+\Delta q\}$. By analysing the CAFs of suspensions P, M1 and M2 as functions of $\phi$ and $q$ we arrive at the (normalised) width, $\Delta q/q_m$, of this window shown in Fig. 3.

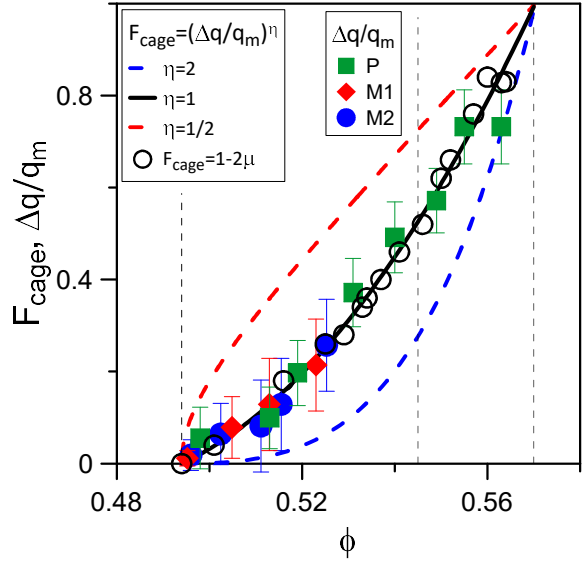

FIG. 3. Filled symbols show $\Delta q/q_m$ determined from the CAFs of suspensions P, M1 and M2 whose decays display variations in curvature. For reference, open circles show $F_{cage}=1-2\mu$, defined in section III.C. Lines, based on the quadratic fit of $\mu$ versus $\phi$ (discussed in Sec. III.D.1), show $(\Delta q/q_m)^\eta$ for values of $\eta$ indicated in the (left) legend. Vertical dashed lines are located at $\phi_f$, $\phi_{1/2}$ and $\phi_g$, as defined in the text.

For those spatial frequencies, $q \neq q^*$, in the complementary window(s), $q \notin \{q_m-\Delta q, q_m+\Delta q\}$, the CAFs decay by the same SE and, therefore, express the *same* Brownian current reversals as those of the thermodynamically equilibrated suspension "at" packing fraction $\phi_f$; ie, $C(q \neq q^*,\tau;\phi>\phi_f)=C(q,\tau;\phi_f)$[48, 50, 51]. The quotation marks here are meant to convey the limitation in the precision to which the freezing point can be identified experimentally[68]. Note, whatever the means of transport, be it by "collisions" for $q=q^*$ or diffusion for $q \neq q^*$, conservation of number density is the overriding constraint that renders the CAF negative in the experimental time window. The difference is that for $q=q^*$ current reversals are correlated, while for $q \neq q^*$ they are not; fluctuations in the current, $j(q,t)$, are non-Gaussian for $q=q^*$ and Gaussian for $q \neq q^*$; interpreting this as an elastic response for $q=q^*$ and viscous flow for $q \neq q^*$. To be clear; for $q \neq q^*$ the longitudinal currents and the motion of the particles comprising those currents are Brownian and, by the persistence of the SE decays of corresponding CAFs to the experimental noise floor, there is no evidence of exchanges between Brownian and caged particles.

Accordingly, we posit that caged particles remain so and, as foreshadowed in Sec. I, delay in decay of time correlation functions is due to correlated cage fluctuations rather than occasional inter-cage exchanges. However, these inferences apply only for the reversible processes under scrutiny here. We discuss irreversible vis-à-vis reversible processes in Sec. III.F after we have quantified the reversible collective dynamics in the ensuing analyses.

Toward that quantification we point out that the values of the exponents, $\nu$, of the power law approximations in Fig. 1b and 2, exceed -2. As we will find in Sec. III.D, the significance of the difference, whether $\nu<-2$ or $\nu>-2$, is that it determines whether correlations of cage fluctuations are propagating or non-propagating; ie, phononic or non-phononic. Or, in the language of the mode-coupling theory (MCT)[45], whether those correlations decay by the critical or the von Schweidler law. Both are encompassed in what is generally referred to as the "β-regime".

**B. Dynamical heterogeneity; a framework for solidification.**

We represent the preceding observations schematically in Fig. 4. Dynamical heterogeneity is expressed by the spatial separation of currents of diffusing particles at packing fraction $\phi_f$, in the window(s) $q \gtrless |(q_m \pm \Delta q)|$, from currents that display caging dynamics among particles at packing fraction $\phi_{solid}$ in the window $q_m \pm \Delta q$. $\phi_{solid}$ remains to be determined.

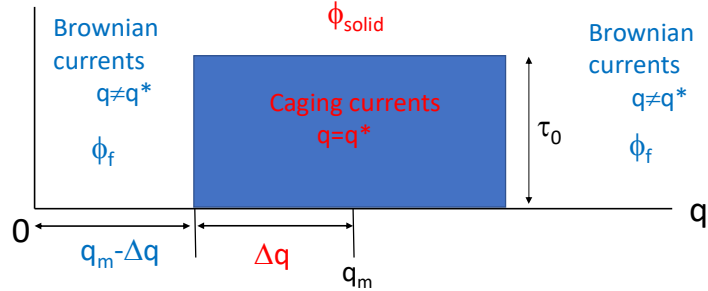

FIG. 4. Dynamical framework for vitrification. The solid obtains in the space-time window $2\Delta q \times \tau_0$. The height $\tau_0$ is defined in Eq. (12).

In this framework we define a dimensionless dynamical length,

$$L_D = q_m/(q_m-\Delta q). \tag{11}$$

We express the slowing dynamics attendant to the increase in $L_D$ with a dimensionless time scale

$$\tau_0 = L_D - 1. \tag{12}$$

Eq. (11) quantifies the dynamical length, introduced in Sec. I, or spatial frequency, $\sim 1/L_D$, that separates dynamics characteristic of an elastic solid from that of a viscous fluid; or simply and for brevity, $L_D$ separates "solid" and "fluid". However, since the Brownian currents, $C(q \neq q^*, \tau)$, remain the same as those observed at $\phi_f$ any increase in delay of structural relaxation attendant on super-packing is due entirely to changes in the collective processes in the solid rather than any increase in viscosity of the fluid. This answers the first question posed in Sec. I.

The solid obtains in the space-time window $2\Delta q \times \tau_0$. What, if any, aspect of caging dynamics of the solid $\tau_0$ describes remains to be determined. For now we note, from Eq. (11) and (12), that both $L_D$ and $\tau_0$ diverge in the limit, $\Delta q \to q_m$, which we identify here with the transition to an ideal glass. At this point there are Brownian currents only for spatial frequencies $q > 2q_m$; ie, localised Brownian currents. Arbitrary subtraction of one in Eq. (12) ensures that in the limit where $\phi_f$ is approached from above and caging vanishes ($\Delta q \to 0$) we have $L_D \to 1$ and $\tau_0 \to 0$.

As illustrated in Fig. 2, structural relaxation in the solid comprises (I) intra-cage rattling and (III) the α-process, separated by a weakly curved crossover (II), $C(q^*, \tau_{1x} < \tau < \tau_{3x})$, approximated by a power-law, $\sim \tau^\nu$. However, as foreshadowed in the close of Sec. III.A this is not the only possible relaxation scenario in the β-regime bridging intra-cage rattling and the α-process. Sec. III.D and III.E are devoted entirely to these scenarios and the manner of their crossover, at $\tau_{3x}$, to the α decay. Before this we determine the fraction, $F_{cage}$, of caged particles.

*Summary A & B*

*We define dynamical heterogeneity by the separation of a dynamical solid of caging currents in the window of spatial frequencies, $q^* \in \{q_m - \Delta q, q_m + \Delta q\}$, from a dynamical fluid of Brownian currents in the complementary window, $q \notin \{q_m - \Delta q, q_m + \Delta q\}$. This separation distinguishes the metastable suspension from the thermodynamically stable suspension which, in answer to the second question posed in Sec. I, is dynamically homogeneous and devoid of caging.*

**C. Determination of the fraction, $F_{cage}$, of caged particles.**

*1. From the MSD by DLS.*

We consider here quantities derived from the self ISF, Eq. (5) to (7). First, Fig. 5a illustrates the increase in stretching of the MSD that typically accompanies an increase in packing fraction[49]. Stretching indices, μ, and root-mean-squared displacements (RMSD), $R_m$, where stretching is greatest are shown in Fig. 5b. Extrapolation of μ to zero (μ→0) gives an estimate of the packing fraction, $\phi_g \approx 0.57$, of the "ideal" glass at which point all particles are presumed to be confined to their respective neighbour cages. Save discussion of irreversible processes in Sec. III.F, this ideality is implicit in further references to the "glass". Between the extremes, $\mu(\phi\to 0)\to 1$ and $\mu(\phi\to\phi_g)\to 0$, of an infinitely dilute suspension of freely diffusing particles and the glass, we see that $\mu(\phi\approx\phi_f)=1/2$; ie, in the super-packed regime ($\phi_f<\phi<\phi_g$) $\mu$ decreases from ½ to zero.

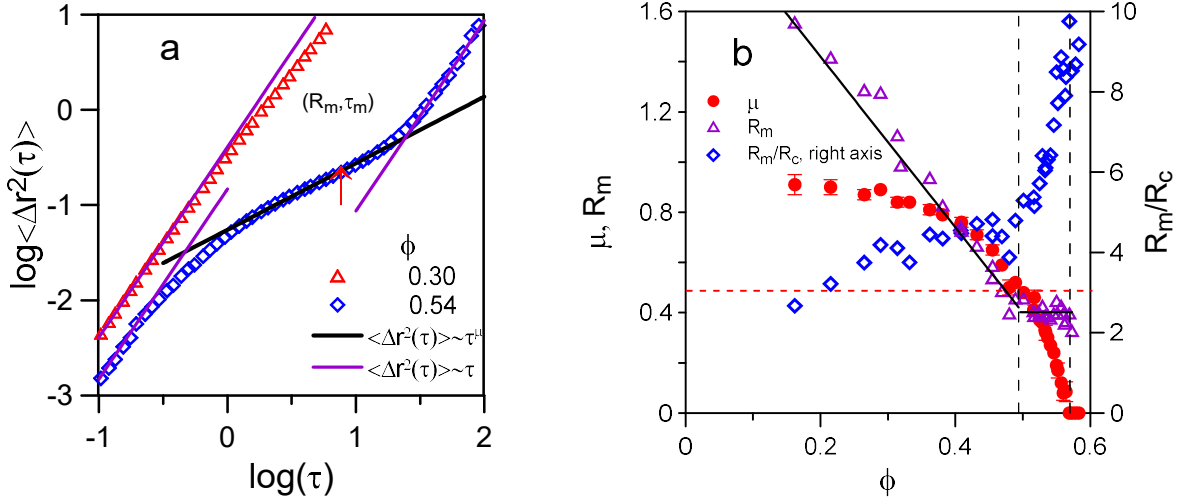

FIG. 5. (a) Double logarithm plot of the mean squared displacements for packing fractions indicated. Magenta lines of unit slope ($\Delta r^2(\tau)\sim\tau$) indicate diffusion. $(R_m,\tau_m)$ locates the point of maximum stretching of the MSD and the black line, of slope $\mu$, is the tangent through it. (b) Stretching index, $\mu$ (Eq. 7), and absolute and relative RMSDs, $R_m$ and $R_m/R_c$, where $R_c=(\phi_R/\phi)^{1/3}-1$ is the average inter-surface spacing and $\phi_R=0.64$ is the packing fraction at random close packing. The dashed vertical lines indicate freezing and glass transition points, $\phi_f$ and $\phi_g$. The solid lines through $R_m$ serve to indicate that it decreases linearly with $\phi$ up to approximately $\phi_f$ and, between $\phi_f$ and $\phi_g$, its value, $R_m\approx 0.4$, shows no systematic variation

By the discussion in Sec.I and the reasoning applied to the CAF in Sec. III.A, we attribute stretching of the MSD of the thermodynamically equilibrated suspension ($\phi<\phi_f$) to the spread in diffusion rates of independently diffusing particles. We then ascribe further stretching of the MSD, evident by reduction in $\mu$ below ½ (Fig. 5b) seen on super-packing ($\phi>\phi_f$), to phase maintaining, correlated reversals along the direction of a particle's motion – the collective dynamical manifestation of caging projected onto the MSD. As more particles in the suspension participate in these correlated reversals so the stretching of the MSD increases. Accordingly, we propose that the fraction of caged particles is given by $F_{cage}=1-2\mu$. The results are plotted in Fig. 6. As expected $F_{cage}$ increases uniformly from zero to one as $\mu$ decreases from ½ to zero. Note also that the point where half the particles are caged apparently coincides with the melting packing fraction ($\phi_m=0.545$) of the one-component hard sphere crystal. However, we cannot attach significance to this since the actual melting volume fractions are slightly reduced by the suspensions' polydispersities (Table I). Accordingly, we label the packing fraction where $F_{cage}=1/2$ as $\phi_{1/2}$.

On the face of it these correlated reversals appear to have no effect locally since, as one sees from Fig. 5b, the RMSD, $R_m(\approx 0.4)$, for $\phi > \phi_f$ shows no systematic variation with $\phi$. However, we find that the ratio, $\delta R = R_m/R_c$, increases from $\delta R(\phi_f) \approx 5$ to $\delta R(\phi_g) \approx 10$. Here $R_c = (\phi_R/\phi)^{1/3} - 1$ is the particles' average inter-surface spacing. We presume here that the particle distributions in the colloidal fluid (at $\phi_f$) and glass (at $\phi_g$) are random so that both are able to be compressed to random close packing, $\phi_R = 0.64$. The upshot is that in the glass, where particles are ostensibly confined to their neighbour cages, there are correlated cage fluctuations allowing RMSDs some 10 times the inter-surface spacing.

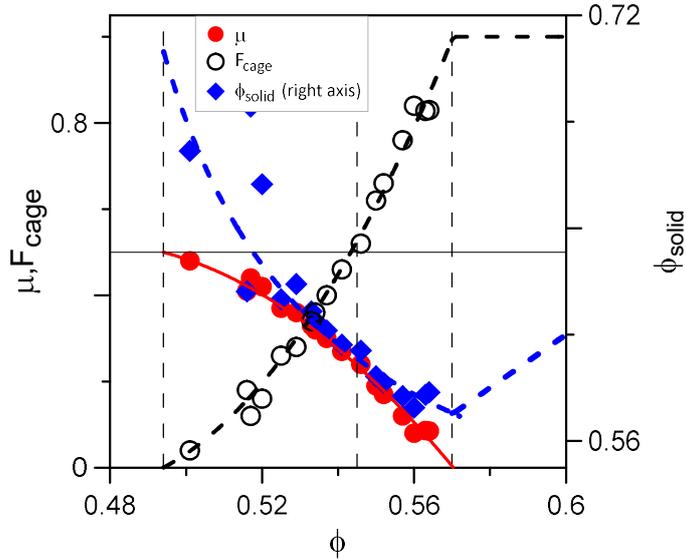

FIG. 6. Properties of the solid obtained from the MSD of suspension T; stretching index, $\mu$ (red discs) and $F_{cage} = 1 - 2\mu$ (open circles). The other property, derived in the text is $\phi_{solid}$ (blue diamonds, Eq (14)). The vertical dashed lines locate points, $\phi_f = 0.494$, $\phi_{1/2} = 0.545$ and $\phi_g = 0.57$. Note, for $\phi_g < \phi < \phi_R$, $\mu = 0$, $F_{cage} = 1$ and $\phi_{solid} = \phi$. In this and subsequent figures symbols express or derive directly from the actual data and curves are based on the quadratic fit to $F_{cage}$. See text in Sec. III.D for details.

*2. From molecular dynamics.*

Stretching indices, $\mu$(MD), obtained by MD[54] of a marginal binary mixture of hard spheres are shown in Fig. 7a. The packing fraction at freezing of this mixture is 0.505. So, as was done for the suspensions (Sec. II), packing fractions are scaled by the factor, 0.494/0.505 in this case, that expresses them relative to the freezing point, $\phi_f = 0.494$, of the one component hard sphere fluid. Then by the reasoning just applied to the suspension, we attribute any reduction in $\mu$ (Fig. 7a) seen for packing fractions beyond $\phi_f$ to caging. We note, on account of the particles' motions being ballistic the stretching indices are larger than those of the suspension; specifically, at $\phi_f$ $\mu$(MD)$\approx 0.8$. However, when the stretching indices from MD are scaled by the factor $\mu$(DLS)/$\mu$(MD)$= 0.5/0.8$ they are seen to coincide with the DLS results, as are the corresponding fractions of caged particles, $F_{cage}(MD)=1-2(0.5/0.8)\mu(MD)$ (Fig. 7b). We infer, from this consistency, that caging and its collective consequences are independent of the microscopic dynamics. Similar conclusions were drawn from some of the earliest computer simulation studies of the GT[41, 93]. Note, in particular that $F_{cage}$ is a convexly increasing function of $\phi$. In Sec. III.D we will find that this curvature has significant implications.

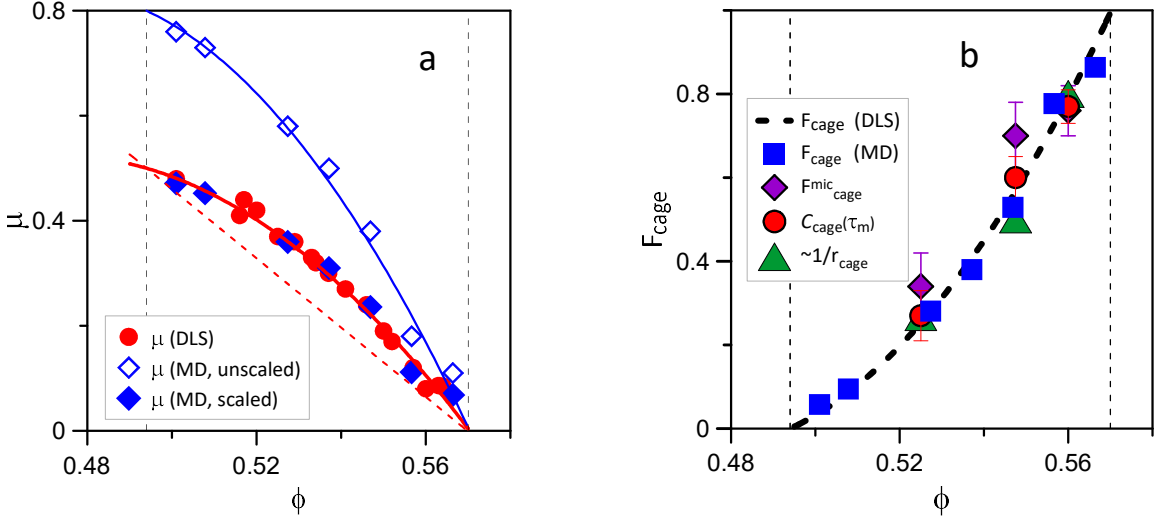

FIG. 7. Comparison of results from molecular dynamics[54], confocal microscopy[23] and DLS. (a) Stretching indices from MD, open diamonds (unscaled), filled diamonds (scaled) and DLS (discs). See text for details. The (red) dashed line serves to highlight the concavity of $\mu(\phi)$. (b) Fraction of caged particles, from MD (blue squares), $F_{cage}^{mic}$ (purple diamonds) and $C_{cage}(\tau_m)$ (red discs) estimated from confocal microscopy and $F_{cage}$ (black dashed line) from DLS. The volume fractions from MD and microscopy are scaled to that of the ideal hard sphere system. The collision rate, $\sim 1/r_{cage}$ (green triangles) determined from microscopy is discussed in Sec. III.D.1. Where error bars are not drawn errors are generally comparable to the size of the symbols. Vertical dashed lines are located at $\phi_f$ and $\phi_g$.

*3. From confocal microscopy.*

Microscopic studies [23, 27] show the rattling of caged particles rather more directly than DLS experiments. We single out the work of Weeks and Weitz[23] since it offers aspects of the cage dynamics that directly complement the above DLS results and, to our knowledge, it gives the only estimate of the fraction of caged particles obtained by confocal microscopy. In their study, Weeks and Weitz measured both the MSD and a topological cage correlation function $C_{cage}(\tau)$, the fraction of particles with the same neighbours at time t and time t+τ averaged over all t. From their data we determine the fraction of caged particles by two routes; First, in the manner employed in the DLS data in Fig. 5a, we determine the delay time, $\tau_m$, where the microscopically determined MSD is most stretched, the corresponding stretching index, μ, and thence the fraction, $F_{cage}^{mic}=1-2\mu$, of caged particles. Second, we read the value of the cage correlation function, $C_{cage}(\tau_m)$, at $\tau_m$. By definition, the latter expresses the fraction of caged particles. To allow meaningful comparison between these estimates, $F_{cage}^{mic}$ and $C_{cage}(\tau_m)$, of the fraction of caged particles and the corresponding quantity, $F_{cage}$, derived from DLS and MD, we map the packing fraction range (0.38, 0.58) between the freezing point and the GT of the suspension studied in Ref. 23 onto that (0.494, 0.57) of our HS suspension. After this re-calibration we see in Fig. 7b that there is no significant inconsistency between the microscopic and spectroscopic estimates of the fraction of caged particles.

Finally, we see from Fig. 3 that the values of $F_{cage}$ overlap with the values of the normalised width, $\Delta q/q_m$, of the spatial window where the occurrence of caging is inferred from the decay of the CAFs (see Sec.III.B). Considering the curves $(\Delta q/q_m)^\eta$ traced for several values of $\eta$ we see, errors notwithstanding, best consistency between experimental results and our framework is obtained for $\eta=1$; ie, $\Delta q/q_m=F_{cage}$. Combining this with Eq. (11) one obtains;

$$F_{cage} = \Delta q/q_m = 1-2\mu = (L_D-1)/L_D \tag{13}$$

*Summary C*

*The fraction of caged particles read from the stretching of the MSD is $F_{cage}=1-2\mu$. In this respect the results from DLS, MD and confocal microscopy are not only consistent with each other but also with the same fraction obtained from the cage correlation function. Moreover, $F_{cage}$ equates with the normalised width, $\Delta q/q_m$, of the (reciprocal) space occupied by the dynamical solid. $F_{cage}$ sets the dynamical length, $L_D$, that quantifies the dynamical heterogeneity.*

**D. Dynamical properties of the solid: From rattling to the β-regime.**

*1. Local rattling.*

In Sec. III.A we determined that there are effectively no reversible exchanges between currents of diffusing particles of the fluid, in the window(s) $q \gtrsim |(q_m \pm \Delta q)|$, and the currents of caged particles of the solid in the window $q_m \pm \Delta q$. Accordingly, the suspension's total/average packing fraction is

$$\phi = F_{cage}\phi_{solid} + (1-F_{cage})\phi_f, \tag{14}$$

where $\phi_f$ and $F_{cage}=1-2\mu$ are given. Evidently, the packing fraction, $\phi_{solid}$, of the solid then obtained, and shown in Fig. 6, *decreases* on increasing super-packing. The implication is that as the fraction of caged particles increases the amorphous solid comprising those particles swells. This seems counterintuitive and, in view of previous estimates of the cage size[23], unexpected. We return to this dilemma below. For now we note that by construction this swelling results from the convexity of $F_{cage}(\phi_f<\phi<\phi_g)$; ie, $d^2F_{cage}(\phi_f<\phi<\phi_g)/d\phi^2>0$. This was noted for both atomic and colloidal hard sphere systems in Sec. III.C.2 and it implies, not implausibly, that caging lags super-packing. However, while the result as such may be robust it suffers error amplification in proportion to $1/F_{cage}$. This merely reflects the fact that little can be inferred about the properties of the infinitesimal fraction of "solid" that first forms when $\phi_f$ is just exceeded. However, by fitting a quadratic, $F_{cage}(\phi)=a(\phi-\phi_f)+b(\phi-\phi_f)^2$ to the data (Fig. 6) the curvature constraint is imposed explicitly and the noise amplification circumvented. Values of the fitting parameters are $a=4.7\pm0.5$ and $b=110\pm10$, and the estimate of the packing fraction of the embryonic solid is $\phi_{solid}(\phi\to\phi_f^{(+)})=\phi_f+1/a=0.71\pm0.02$.

On the face of it swelling of the solid as it develops on super-packing implies a concomitant *increase* of the particles' average inter-surface distance. This would appear to be at odds with Ref. 23 which determines that the cage size *decreases* with increasing $\phi$. However, the latter determination is based on direct observation of the average distance particles travel between reversals in direction of their motion and, as such, it expresses the mean free path rather than the mean inter-surface distance. Accordingly, the reduction in the mean free path, denoted as $r_{cage}$ in Ref. 23, reflects a corresponding *decrease* in collision time $\tau_{coll}$. The inverse $1/r_{cage}$, scaled to coincide with other results, is shown in Fig. 7b.

Paucity of data and errors notwithstanding we submit that the collision rate between caged particles increases in proportion to the fraction of caged particles; $1/\tau_{coll}\sim F_{cage}$. Again we arrive at the implication that there are energy exchanges among those particles wherever their putative Brownian motion is impaired by proximity of neighbouring particles. In this sense $F_{cage}$ sets a basic time scale, $\tau_{coll}\sim 1/F_{cage}$, that then determines the time scales of all collective consequences of the collisions.

With this in mind we interpret the results of Fig. 6 as follows: On super-packing solidification starts with an infinitesimally small amount of solid osmotically compressed, by the longitudinal Brownian currents of the fluid, to structures of average packing fraction $\phi_{solid}(\phi_f)\approx 0.7$. The latter is too large for the structure of this embryonic solid to be randomly close packed ($\phi_R=0.64$). It is not inconsistent, however, with that of compact low order polyhedra nor, for that matter, close-packed crystals too small and few to see[94]. On further super-packing the internal pressure, that builds up on account of the increasing collision rate among the particles in the solid soon exceeds the fluid's osmotic pressure, causing it to swell. We see from Fig 6 that the solid's packing fraction then decreases from a maximum, $\phi_{solid}(\phi_f)\approx 0.7$, to a minimum, $\phi_{solid}(\phi_g)=\phi_g\approx 0.57$, at the GT. Beyond that we have $F_{cage}=1$ and $\phi_{solid}(\phi_g<\phi<\phi_R=0.64)=\phi$.

Having studied the local rattling of the particles of the dynamical solid we next consider the collective consequences exposed in the β-regime.

*2. The β-regime.*

Fig. 8 shows the CAFs for a range of packing fractions spanning the metastable fluid and glass-like suspensions. Decays that approximate to power laws, $\sim\tau^\nu$, are evident in all cases. Exponents, $\nu$, are shown in Fig. 9a (left axis). We drop the q dependence from the various parameters below on the understanding these results and their discussion apply for the spatial frequency $q_m$.

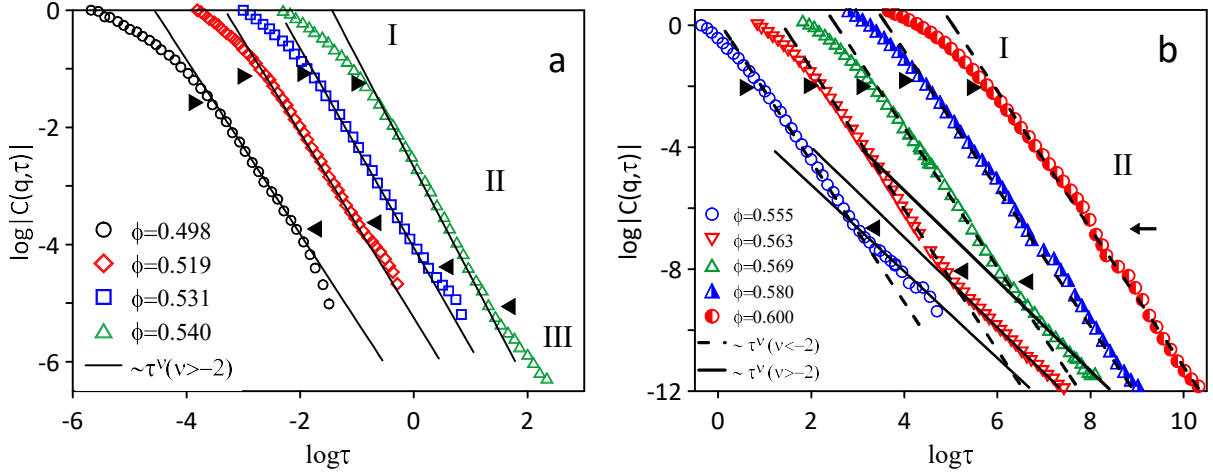

FIG. 8. CAFs, $C(q_m,\tau)$ for packing fractions as indicated in (a) $\phi_f < \phi < \phi_{1/2}$ and (b) $\phi_{1/2} < \phi < \phi_R$. Data are shifted on the $\log\tau$-axis for clarity. Note the difference in the scales of axes in (a) and (b). Power laws, $\sim\tau^\nu$, are shown as continuous ($\nu > -2$) and dashed ($\nu < -2$) lines which in the discussion we interpret, respectively, as non-propagating and propagating cage fluctuations. In (a) symbols, ▶ and ◀, indicate approximately the crossover times $\tau_{1x}$ and $\tau_{3x}$, respectively. In (b) symbols, ▶ and ◀, indicate crossover times $\tau_{1x}$ and $\tau_{2x}$, respectively. See Table II for definitions. For $\phi_{1/2} < \phi < \phi_g$ no α-process can be identified in the CAF and a second power law (continuous line) has been fitted to what appears as the final decay in the experimental time window. In the glass ($\phi > \phi_g$) there is just one power law. Here the arrow indicates the region where the results of different DLS detection methods overlap[74].

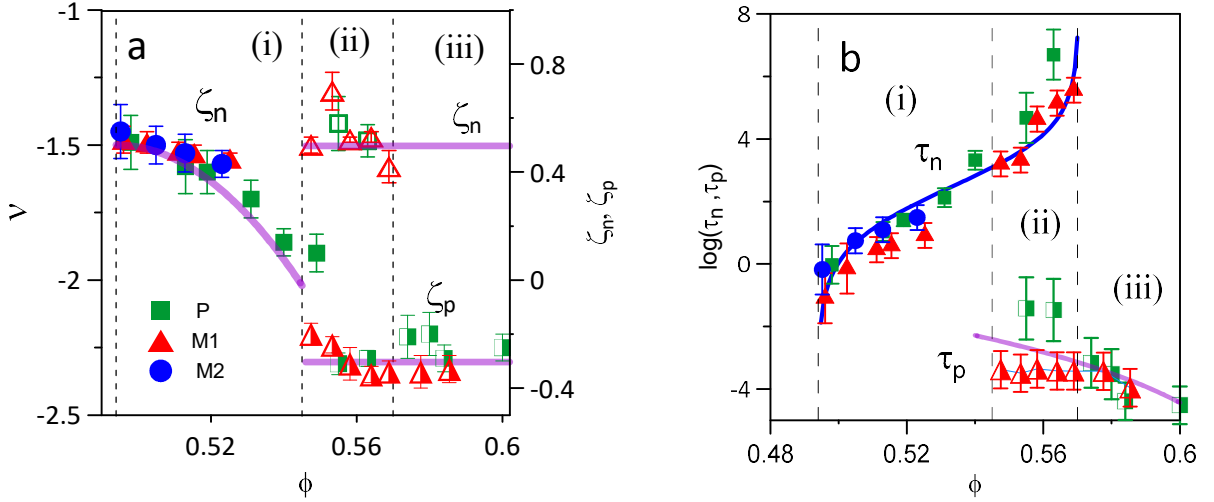

FIG. 9. (a) Exponents, ν, of power laws fitted to the CAFs (left axis). The exponents ($\zeta_{n,p}=\nu+2$) of the corresponding power laws in the ISFs (right axis) are $\zeta_n$ (for ν>-2) and $\zeta_p$ (for ν<-2). Filled symbols are used, when there is just one power law, open and half-filled symbols when two power laws are present. The remaining features of this figure are explained in discussion of the ISF. (b) Power-law time scales $\tau_n$ (filled symbols, Eq. (15) and (16)) and $\tau_p$ (half-filled symbols, Eq. (15) and (17)). Subscripts "n" and "p" refer to non-propagating and propagating modes, respectively. Thick purple lines serve to indicate trends through the data. The solid blue line in (b) traced through the data scales as $\tau_0^2$ (Eq. (12)). Vertical dashed lines delineate the three scenarios of the β-regime; (i) $\phi_f<\phi<\phi_{1/2}$, (ii) $\phi_{1/2}<\phi<\phi_g$, (iii) $\phi>\phi_g$.

By means of these algebraic decays we identify three dynamical scenarios in the β-regime:

(i) In what we will refer to as the "weakly super-packed" region, $\phi_f<\phi\lesssim\phi_{1/2}$ (Fig. 8a), the β-regime is identified by one power law, $\sim\tau^\nu$. Deviation from that power law at long times ($\tau\gtrsim\tau_{3x}$) is how we identify the α-process. Its exposure in the CAF, for reasons discussed in Sec, III.A, is feeble and cannot be quantified with confidence. The power-law exponent (Fig. 9a) decreases more or less uniformly from $\nu(\phi\approx\phi_f)\approx-1.5$ to $\nu(\phi\approx\phi_{1/2})\approx-2$.

(ii) In the "strongly super-packed" region, $\phi_{1/2}\lesssim\phi<\phi_g$ (Fig. 8b), where more than half the particles are caged, we find two successive power laws, one crossing over to the other at $\tau_{2x}$. The second, weaker power law takes us to the noise floor and the crossover to the α-process is not seen in the experimental time window. Here the exponents, ν≈-2.3 and ν≈-1.5 (Fig. 9a), of the successive power laws show no systematic dependence on ϕ.

(iii) Then, in the colloidal glass, $\phi>\phi_g$ (Fig. 8b), following intra-cage rattling (I) just the one power law with ν≈-2.3 approximates the decay of the CAF to the noise floor.

Now we note that power laws, $a\tau^\nu$, in the CAF derive from power law contributions, $a\tau^{\nu+2}/((\nu+2)(\nu+1))$, in the ISF (Eq. (3)). Accordingly, we identify power laws with exponents ν>-2 and ν<-2 in the CAF with concave and convex decays, $-(\tau/\tau_n)^{\zeta_n}$ and $(\tau/\tau_p)^{\zeta_p}$, respectively, in the ISF. The exponents, $\zeta_n$ (=ν+2>0) and $\zeta_p$ (=ν+2<0), are shown in Fig. 9a (right axis), and the time scales,

$$\tau_n=[a_n/|\zeta_n(\zeta_n-1)|]^{-1/\zeta_n} \text{ and } \tau_p=[a_p/|\zeta_p(\zeta_p-1)|]^{-1/\zeta_p} \qquad (15)$$

in Fig. 9b.

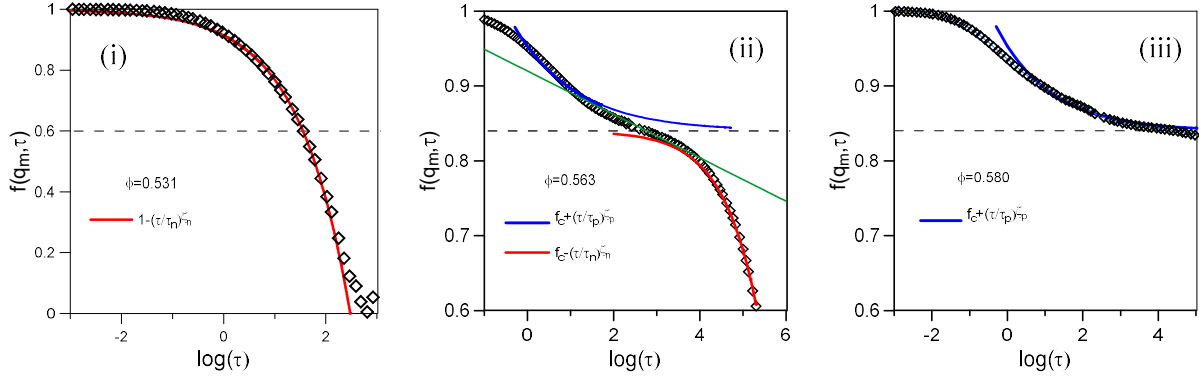

FIG. 10. Experimental ISFs (open diamonds) for packing fractions indicated and spatial frequency $q_m$. Concave (red lines) and convex (blue lines) power laws are traced for average values of the exponents and time scales in Fig. 9. The three scenarios of the β-regime are respectively illustrated in (i) $\phi_f<\phi<\phi_{1/2}$, (ii) $\phi_{1/2}<\phi<\phi_g$, (iii) $\phi>\phi_g$. In (ii) and (iii) $f_c=f_c(q_m)=0.84$ is the non-ergodicity value. In (ii) the green line is the tangent to $f(q_m,\tau)$ through the crossover $\tau_{2x}$. See text for details.

So, having relied up to this point on the CAFs to identify algebraic decays in the suspension's dynamics we proceed discussion in terms of the more familiar ISFs. These are shown in Fig. 10 for cases illustrative of the three relaxation scenarios of the β-regime. Also shown are the convex and concave power laws just defined, the distinction between them now obvious. Mechanisms we proffer for these power laws are guided, firstly, by the fact that elasticity of supercooled fluids and glasses is due their ability to transmit transverse stresses [32, 33, 36]. However, in DLS experiments we observe the statistics of the responding strain fluctuations projected on the propagation vector, **q**. With this in mind, we attribute the convex and concave power-law decays in the ISF to propagating/underdamped and non-propagating/overdamped correlated cage fluctuations, as indicated by the subscripts, p and n, on the respective power law parameters. Secondly, as mentioned in Sec. I, collective processes occur intermittently[22, 37-39]. Ergo, the power law decays represent non-stationary, intermittent, rather than stationary, continuous, processes. The exponents, $\zeta_n$ or $\zeta_p$, being positive or negative, express the increase or decrease in some length scale associated with the non-stationary process. And the time scales, $\tau_n$ and $\tau_p$, express the interval statistics – the mean waiting times between successive events; their inverses express average intermittency rates.

In light of these connections, we discuss the three scenarios in the β-regime displayed by the ISFs in Fig. 10:

(i) For $\phi_f < \phi \lesssim \phi_{1/2}$ the β-regime (Fig. 10 (i)) is identified by the concave power law, $\sim -(\tau/\tau_n)^{\zeta_n}$, expressing non-stationary, non-propagating cage fluctuations. In the context of our solidification framework (Fig. 4) we express the extent of their spatial correlation at delay time $\tau$ as $q_m \pm \delta q(\tau)$. So, as that spatial correlation grows from $\delta q(\tau_{1x}) \approx 0$ to the limit $\delta q(\tau_{3x}) = \Delta q$, the decay of the ISF can be expressed as

$$f_n^{(i)}(\tau_{1x} < \tau < \tau_{3x}) = f(\tau_{1x}) - (\tau/\tau_n)^{\zeta_n} \qquad . \qquad (16)$$

Here $\tau_{1x}$ is the delay time when that correlation can be first discerned from local rattling (see Fig. 2) and $\tau_{3x}$ is the delay time when they span the reciprocal space, $q_m \pm \Delta q$, occupied by the solid. At this point they become stationary and, as a consequence, the decays of the time correlation functions, be that the ISF or CAF, start to deviate from power laws. In the discussion of Fig. 8a above we attributed the onset of these deviations, however weak and difficult to quantify, to the crossover to the α-process. Note again, this crossover is a consequence of the limit of the space, $q_m \pm \Delta q$, occupied by the solid; a consequence of the dynamical heterogeneity in the super-packed suspension. We discuss the α-process in more detail in Sec. III.E.

We see from Fig. 9 that the exponent of these non-propagating modes decreases from $\zeta_n(\phi \approx \phi_f) \approx 0.5$ to $\zeta_n(\phi \approx \phi_{1/2}) \approx 0$, and time scale, $\tau_n$, increases. Errors aside, the interpretation we offer is that as $\phi \to \phi_{1/2}$ and $\zeta_n \to 0$ a plateau develops in the ISF. Then in proximity of $\phi_{1/2}$, the point where half the particles are caged, the dynamics changes to another scenario;

(ii) $\phi_{1/2} \lesssim \phi < \phi_g$, where we identify successive convex, $\sim (\tau/\tau_p)^{\zeta_p}$, and concave, $\sim -(\tau/\tau_n)^{\zeta_n}$, power laws in the β-regime. Neither exponent, $\zeta_p \approx -0.3$ or $\zeta_n \approx 0.5$, shows systematic dependence on $\phi$ (Fig. 9a). Of itself the convex power law necessarily asymptotes to a plateau, $f_c$. Accordingly, write its contribution to the ISF as,

$$f_p(\tau) = f_c + (\tau/\tau_p)^{\zeta_p} \qquad . \qquad (17)$$

This then expresses the decrease in amplitude of propagating modes, as their wavelength, or spatial correlation, increases from 1 to $(1-F_{cage})^{-1}$ (Eq. (13)). Note that the time scale, $\tau_p$ ($\lesssim 10^{-2}$), in this case is more than two decades *smaller* than the Brownian time and, were one to ascribe a trend to the spread in the results in Fig. 9b, *decreases* with $\phi$. The propagating modes are supported by a structure, $f_c$, that percolates the solid. We identify $f_c$ with the non-ergodicity value[2, 3, 95] at $q_m$. From that structure the concave power law

$$f_n^{(ii)}(\tau) = f_c - (\tau/\tau_n)^{\zeta_n}, \qquad (18)$$

expresses the decay of the ISF by the non-propagating, non-stationary cage fluctuations we encountered under scenario (i). As noted already, the exponent, $\zeta_n \approx 0.5$, shows no systematic dependence on $\phi$. The time scale, $\tau_n$, continues its increase with $\phi$ (Fig. 9b). Also, since $F_{cage}=\Delta q/q_m$ (Fig. 6) is less than one this process, like that in (i), becomes stationary when its spatial correlation spans the space, $q_m \pm \Delta q$, occupied by the solid. However, the non-stationary to stationary cross-over, at $\tau_{3x}$, is beneath the noise floor in the CAF and, due to the ambiguity in identifying power-laws (Sec. I), it is also difficult to pinpoint in the ISF.

Insofar as the exponents, $\zeta_p$ and $\zeta_n$, show no systematic dependence on $\phi$ there is another scale invariance in the dynamics when the fraction of particles exceeds ½. That is, the non-stationarities of the processes expressed by Eq. (17) and (18) are inherent in the dynamics of the solid that obtains in the strongly super-packed suspension and independent of the space, $q_m \pm \Delta q$, that solid occupies. That space, as just seen, limits the spatial correlation of non-propagating cage fluctuations described by Eq. (18) and, by Eq. (11), places an upper limit on the wavelength, $\sim L_D$, of propagating cage fluctuations described by Eq. (17). Furthermore, the relationship,

$$\tau_{2x}^{(\zeta_n-\zeta_p)} \sim \tau_p^{-\zeta_p} \tau_n^{\zeta_n}, \qquad (19)$$

between the time scales of the convex and concave power-laws, obtained from their crossover at $\tau_{2x}$, illustrates the essential intrinsic connection between collective processes that ultimately stem from the same deterministic rattling of the particles (see Sec. III.D.1).

One sees from Fig. 9b that the time scale $\tau_n$ scales with $\phi$ approximately as $\tau_0^2$ (Eq. (12)) and diverges as $\phi \rightarrow \phi_g$. Here we cross over to,

(iii) the dynamics of an ideal glass ($\phi > \phi_g$) where just the convex power law decay to the plateau (Eq. (17)) remains.

By the scale invariance expressed by the power laws of the β-regime we recognise, as other authors[22, 25, 96, 97] have from different aspects of the GT phenomenology, characteristics of self-organised criticality common to many natural processes[98].

*Summary D*

*The collision rate between caged particles increases in proportion to the fraction of caged particles: $1/\tau_{coll} \sim F_{cage}$. Power-law decays express the first identifiable collective consequences of these collisions – correlated cage fluctuations by which we identify the β-regime. These collective modes are intermittent and non-stationary and they can be either propagating or non-propagating. We identify three scenarios: (i) For weak super-packing ($F_{cage}<1/2$) the β-regime comprises non-propagating modes; (ii) For strong super-packing ($1/2<F_{cage}<1$), there are propagating and non-propagating modes, manifested in the ISFs by decays to and from a plateau; (iii) For the ideal glass ($F_{cage}=1$) only propagating modes remain and these effect the decay to the plateau.*

*In (i) and (ii) the spatial correlation of non-propagating modes spreads with delay time until it fills the space, $q_m \pm \Delta q$, occupied by the solid. They then cross over from non-stationary to stationary which we identify with the crossover of the β-regime to the α-process.*

### E. The α-process.

Since the α process applies only for the super-packed suspension the decay times, $\tau_\alpha$, shown in Fig. 11, have been estimated from $\tau_\alpha(\phi) = \tau_{SE}(\phi) - \tau_{SE}(\phi_f)$, where $\tau_{SE}(\phi)$ is the decay time of the SE (Eq. (8)) fitted to the final decays of the ISF (Eq. (1)) and self ISF (Eq. (5)). The effect of the subtraction $\tau_{SE}(\phi_f)$ is negligible for appreciable super-packing but, by the same token, exposes increasingly large errors when, on approaching $\phi_f$ from above, the decreasing amount of "solid" render its properties increasingly imprecise. We first encountered this limitation in Sec. III.D.1 in our estimation of the packing fraction, $\phi_{solid}(\phi \to \phi_f^{(+)})$, of the solid for the very weakly super-packed suspension.

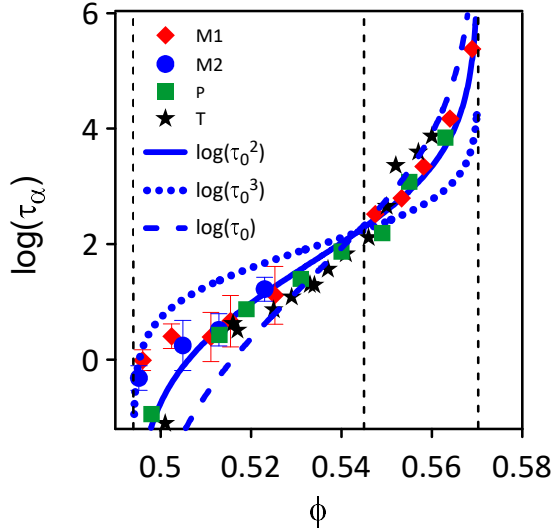

FIG. 11. Symbols show structural relaxation times, $\tau_\alpha$, determined from SE fits to the ISFs measured at $q_m$ for suspensions listed in Table I. (See text for a more detailed description of different procedures for estimating $\tau_\alpha$.) Results for suspension T have been scaled to overlap with the other data. Lines, $\sim\tau_0$, $\sim\tau_0^2$ and $\sim\tau_0^3$, discussed in the text are also drawn. The vertical dashed lines are located at $\phi_f$, $\phi_{1/2}$, and $\phi_g$.

Fig. 11 also plots $\tau_0^\eta$ for several values of η. The time scale $\tau_0$ is defined in Eq. (12). Since $\tau_0^2$ is the most consistent with experimental estimates of $\tau_\alpha$ we propose that the decay time of the α process scales as

$$\tau_\alpha \sim \tau_0^2 \sim [F_{cage}/(1-F_{cage})]^2 \sim (L_D - 1)^2 \qquad (20)$$

So, as $\phi \rightarrow \phi_g$, $\tau_\alpha$ scales as $L_D^2$. Thus, while the present data does not preclude a fractional exponent of $L_D$, the α-process appears closer to areal than either linear or volumetric, in space. So, as surmised in Sec. III.D.2, collective processes observed by DLS, of which the α process is the slowest, express the longitudinal projection of (transverse) strain fluctuations.

We found in Sec. III.D that $\tau_{3x}$ locates the crossover from the β-regime (II) to the α-process (III) or, more precisely, the termination of the concave power-law in the ISF that marks the crossover of the non-propagating cage fluctuations from non-stationary to stationary. Thus, the α process is an intermittent stationary process. Further to this we note that the time scale, $\tau_n$ (Fig. 9b), of the non-stationary modes, from their onset at $\phi_f$ to their arrest at $\phi_g$, scales approximately with $\tau_0^2$ in the same manner as $\tau_\alpha$. That is, the β- and α-processes share the same interval statistics, as expressed by the time scale $\tau_n$.

Finally, we must clarify that by application of SE approximations to estimate $\tau_\alpha$ we do not imply that the α process comprises a sum of random, Gaussian processes.

*Summary E*

*The α-process expresses final structural relaxation by the de-phasing of the density field through stationary, intermittent collective particle motions in the solid rather than, as is perhaps more usually pictured (Sec. I), through melting or decay of the amorphous structure per se. Neither power law nor stretched exponential rigorously characterises the α process.*

**F. Irreversible processes; aging and crystallisation.**

Reversible and irreversible processes generally comprise interdependent dynamics in super-packed/super-cooled fluids. However, as discussed in Sec. I, II and III.B, with the benefits of a nearly symmetrical particle size distribution of moderate width and our focus on the CAF we have effectively exposed just the reversible processes. Their action alone maintains the integrity of the amorphous solid – the dynamical heterogeneity of the super-packed suspension – and exposes an ideal/reference vitrification mechanism. All its characteristics are determined by $F_{cage}(\phi)$ which, by Eq. (14), also determines $\phi_{solid}(\phi)$. For the purpose of this discussion, both quantities are retraced in Fig. 12. Accordingly, we submit that exchanges of caged particles either with other caged particles in the solid or with diffusing particles in the fluid – exchanges that would disrupt the integrity of the solid – are irreversible and not accounted for in the curves, $F_{cage}(\phi)$ and $\phi_{solid}(\phi)$.

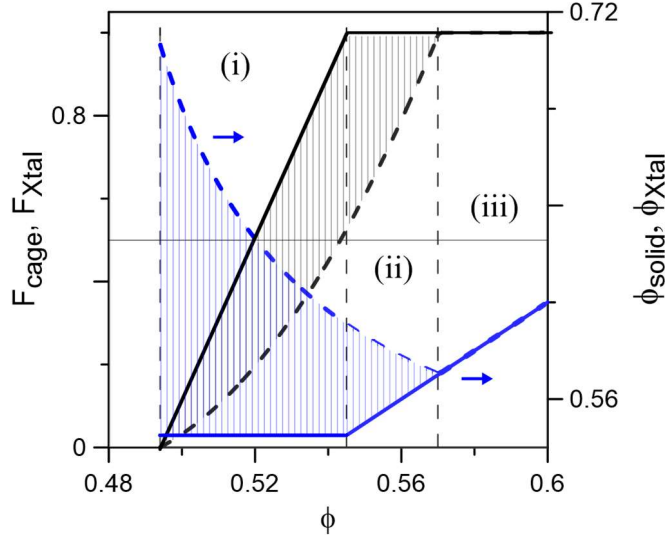

FIG. 12. Schematic comparison of properties, $F_{cage}$ and $\phi_{solid}$, of the amorphous solid (dashed lines) and the corresponding properties, $F_{Xtal}$ and $\phi_{Xtal}$, of the equilibrium crystalline solid. Note that $\phi_{solid}$ and $\phi_{Xtal}$, are referred to the right axis. From the HS phase diagram we have $\phi_{Xtal}(\phi_f<\phi<\phi_m)=\phi_m$ and $\phi_{Xtal}(\phi>\phi_m)=\phi$. And by applying a lever rule, $F_{Xtal}(\phi_f<\phi<\phi_m)=(\phi-\phi_f)/(\phi_m-\phi_f)$, and $F_{Xtal}(\phi>\phi_m)=1$. Dashed vertical lines, located at $\phi_f$, $\phi_{1/2}$ (=$\phi_m$ in this schematic diagram) and $\phi_g$, separate the three dynamical scenarios of the β regime.

Now we see from Fig.12 that departures from those curves must proceed along vertical lines of fixed suspension packing fraction ϕ, in the direction of the corresponding quantities $F_{xtal}(\phi)$ and $\phi_{xtal}(\phi)$ that characterise the equilibrium crystal.

Before discussing the implications of this, we consider relevant experimental findings. First, during the early aging stages of strongly super-packed suspensions DLS measurements of the ISF and CAF[53, 74] find that the plateau value, $f_c$ (Eq. (17)), and the power-law decay to that plateau do not depend on the waiting time, $t_w$. But the cross-over time, $\tau_{2x}$, from that power-law to the decay from the plateau increases with $t_w$. The point illustrated here is for "small" departures from the lines $F_{cage}(\phi)$ and $\phi_{solid}(\phi)$, the dynamics are evidently still dominated by the propagating modes in an *amorphous* solid and not by the fraction of the volume that solid occupies in the super-packed suspension. Thus, a "small" irreversible increase in caged particles still maintains the integrity of the amorphous solid comprising those particles.

Second, studies of the crystallisation kinetics [52, 53, 99-102] find after the initial density quench – before emergence of reflections characteristic of Bragg reflecting colloidal crystals[103] – that the area under the main structure factor peak increases, and its position, $q_m$, decreases. So, it appears there's an induction phase during which the amount of solid increases, that solid swells while, at the same time, its structure remains amorphous.

Now we see from Fig. 12 that proceeding from $F_{cage}(\phi)$ and $\phi_{solid}(\phi)$ along vertical lines of fixed ϕ toward $F_{Xtal}(\phi)$ and $\phi_{Xtal}(\phi)$ necessarily incurs an *increase* in the fraction of caged particles – by incorporation of diffusing particles of the fluid into the solid of caged particles – and a concomitant *decrease* in the packing fraction of that solid. These inferences are consistent with the observations just mentioned. We consider this aspect of the irreversible dynamics as early stage aging – where the solid remains essentially amorphous and its swelling is, as asserted for the reversible dynamics in Sec. III.D.1, due to the increase rattling rate among the caged particles consequent on the increase of their fraction in the suspension.

During this (irreversible) swelling of the solid the packing fraction of the residual colloidal fluid increases beyond its value, $\phi_f$, that obtains for the reversible dynamics. The osmotic pressure the resulting over-packed fluid exerts on the solid increases, and increasingly mitigates the growth and swelling of the solid. Thus, the irreversible caging of particles cannot continue indefinitely without change to a more efficiently packed structure; we can but speculate this would involve structural change from amorphous to crystalline. We also surmise that some combination of the differences, $\Delta\phi = \phi_{solid}(\phi) - \phi_{Xtal}(\phi)$ and $\Delta F = F_{Xtal}(\phi) - F_{cage}(\phi)$, factors into the stability of the super-packed suspension to crystallisation. The "ideal" glass would then be unstable to crystallisation

In brief, we infer as follows; Formation of the crystal phase occurs through collective irreversible structural re-arrangements in regions of amorphous solid defined by the dynamical heterogeneity (Fig. 4). These structural re-arrangements are triggered by the very reversible processes – namely, those processes of the β regime – that maintain the integrity of the amorphous solid. These inferences are consistent results from light scattering[100, 101] and computer simulations[102] studies that also find that crystals evolve from amorphous precursors. We would only add, on account of the scale invariant dynamics (Sec. III.D), that the amorphous precursors are in dynamically critical states.

We are aware this account of the separation of the crystal phase is at odds with classical nucleation theory (CNT), originally developed for droplet condensation[104, 105]. Only later, in the absence of an alternative, was this same theory applied to crystallization[106] which, by all accounts, has held sway ever since[107]. Our account of the process is contentious and clearly warrants scrutiny beyond the scope of this paper.

## IV. OUTSTANDING ISSUES

Aside from determination of the spatial window, $q_m \pm \Delta q$, of the dynamical solid, discussion of the dynamics of the super-packed suspension has been generally limited to the spatial frequency, $q_m$. Detailed examination of the dynamics for other spatial frequencies in this window is warranted and will be presented in a future communication.

Among the issues that may be raised we mention two.

The first concerns the source of the deterministic "rattling" from which all subsequent collective processes stem. There is ample evidence showing that in the presence of obstacles, such as a duct[108], a wall[109] or caged particles, a liquid's transverse momentum current propagates. In these cases momentum is not conserved locally and the classical positive, $(+)\tau^{-3/2}$, algebraic decay of the velocity auto-correlation function, characterising that conservation in a quiescent fluid in thermodynamic equilibrium, is cut off by a negative algebraic decay[108-111]. Presumably, caged particles in suspension have a similar effect on the transverse momentum currents in the solvent and their propagation drives the particles' rattling. Judging by the negative algebraic decays of their VAFs a similar mechanism applies for super-cooled atomic fluids[54, 111].

Second, we find that the dynamics of the strongly super-packed suspension ($F_{cage}$>1/2) are governed by two interdependent, non-stationary intermittent processes. In their expressions by Eq. (17) and (18) we recognise the critical decay and von Schweidler law predicted by the idealised MCT of the GT[45]. Moreover, values of the respective exponents, $\zeta_p \approx -0.3$ and $\zeta_n \approx 0.5$, and the scaling properties that follow are consistent with those predicted by that theory. Previous analyses of time correlation functions obtained from computer simulations on simple atomic fluids[112, 113] and DLS experiments on hard sphere suspensions[75, 86] in terms of the MCT have generally required several fitting parameters. It is noteworthy, therefore, that the present predictions of the MCT have been obtained from experimental observations without recourse to that theory! This adds pertinence to the question of how the theory, with its attendant simplifications and approximations, among them the neglect of transverse currents, evidently captures the very basic collective aspects of the observed dynamics?

We defer more detailed discussion of both questions to future publications.

## V. SUMMARY AND CONCLUSIONS

We have examined the dynamics of the hard sphere system with the aim of understanding and connecting various known aspects of the vitrification process. First and foremost, the thermodynamic freezing point of the hard sphere fluid at the packing fraction, $\phi_f$=0.494, delineates the equilibrium ($\phi<\phi_f$) and super-packed ($\phi>\phi_f$) regimes. In the equilibrium case there is no caging and the dynamics is homogeneous. Caging then is the hallmark of the super-packed fluid. All collective manifestations of caging, outlined in Sec. I and identified and quantified in Sec. III of our study, can be linked to the fraction, $F_{cage}$, of caged particles. This connection stems from the collective consequences of the local rattling adopted by the caged particles. Increasing the super-packing, read increasing $F_{cage}$, increases their rattling rate which then serves to increase the efficacy of the collective processes – the non-stationary dynamics of the β-regime – in maintaining the integrity of amorphous regions of caged particles and the dynamical heterogeneity of the super-packed fluid. Concomitantly, de-phasing of the particle number density and flux is delayed. In more detail we find as follows;

1. The equality, $F_{cage}=\Delta q/q_m$, between the fraction of particles that is caged and the width, $\Delta q$, of the spatial window that manifests caging dynamics (Fig. 3) is one of the pivotal inferences of this paper. This window defines the space, $q^*\in\{q_m-\Delta q, q_m+\Delta q\}$, of spatial frequencies where dynamics symptomatic of an elastic solid are evident. In the complementary window, $q\notin\{q_m-\Delta q, q_m+\Delta q\}$, the dynamics corresponds to that of the equilibrium fluid at the packing fraction $\phi_f$. By this spatial separation of an amorphous solid of caged particles and fluid of freely diffusing particles, $F_{cage}$ defines dynamical heterogeneity. At the same time $F_{cage}$ sets the maximum wavelength, $\sim L_D=q_m/(q_m-\Delta q)$, for which the super-packed suspension is elastic and, in the same proportion, $F_{cage}$ sets the fundamental time scale for the decay of the time correlation functions. The limit $F_{cage}\to 1$, $\Delta q\to q_m$, locates the ideal glass, at $\phi_g$, where all particles are caged and the dynamical length and time scales diverge.

2. $F_{cage}$ also determines the packing fraction, $\phi_{solid}$, of the amorphous solid, which, by Eq. (14), decreases with super-packing (Fig. 6). This swelling of the solid is due to the increase in the rattling rate, $1/\tau_{coll}$, which, we contend, occurs in proportion to $F_{cage}$.

3. For weak super-packing, defined by $\phi_f<\phi<\phi_{1/2}$ and $0<F_{cage}<1/2$, the immediate collective manifestation of the rattling is evident by a concave ($\sim -(\tau/\tau_n)^{\zeta_n}$ ($\zeta_n>0$)), power law in the ISF (Fig. 10a), we identify with intermittent, non-stationary, non-propagating cage fluctuations. As $\phi\to\phi_{1/2}$ and concurrently $F_{cage}\to 1/2$, the exponent $\zeta_n$ decreases from approximately ½ to 0.

4. In the strongly super-packed regime, where more than half the particles are caged, $F_{cage}>1/2$, the above concave decay is preceded by convex power law, ($\sim(\tau/\tau_p)^{\zeta_p}$ ($\zeta_p<0$)) in the ISF (Fig. 10b), identified with intermittent, non-stationary but propagating cage fluctuations. These effect the decay of the ISF to a plateau; a percolating structure in the solid, supporting propagating modes that further maintain the stability of the amorphous solid.

5. Insofar as the exponents, $\zeta_p\approx -0.3$ and $\zeta_n\approx 0.5$, show no systematic dependence on $\phi$, there is a scale invariance of the propagating and non-propagating modes of the β-regime of the strongly super-packed suspension. The time scale, $\sim\tau_p$, of the propagating process *decreases* with super-packing. In contrast, the time scale, $\sim\tau_n$, of the non-propagating process *increases* and diverges as $\phi\to\phi_g$.

6. In all cases ($\phi<\phi_g$) the spatial correlation of the non-propagating modes spreads with increasing delay time until that correlation encounters a limit given by the spatial extent, $q_m\pm\Delta q$, of the amorphous solid. These modes then cross over from non-stationary to stationary. This we identify with the crossover from the β-regime to the α-process. The time scale, $\tau_n$, that characterises the intermittency of the non-propagating process, and the time scale, $\tau_\alpha$, of the alpha decay both scale as $(1-F_{cage})^{-2}$. Both diverge as $F_{cage} \rightarrow 1$ and $\phi \rightarrow \phi_g$, from which point just the propagating modes remain.

The preceding summarises the characteristics of the reversible processes, those processes that maintain stability of the amorphous solid component in the dynamically heterogeneous, super-packed suspension. From this we are able to construct a schematic (Fig. 12) that indicates constraints on the irreversible processes that would lead to the separation of the crystalline solid.

We close with the following; The power-law decays of the β-regime express the quintessential emergent dynamics that stem from amorphous assemblies of rattling particles. They share their scale invariance – intermittence, non-stationarity – with many naturally occurring, non-equilibrium phenomena[97, 98]. And, one might say, by definition, emergent properties are not evident, cannot be predicted for that matter, from microscopic observations of individual particles.

| Time | Definition | Comments |
| --- | --- | --- |
| $\tau_0$ | $F_{cage}/(1-F_{cage})$, Eq. (12) | Fundamental time characterising dynamical heterogeneity |
| $\tau_{1x}$ | First crossover, from rattling to correlated cage fluctuations | |
| $\tau_{2x}$ | Crossover from propagating to non-propagating cage fluctuations. | Crossover from critical decay to the von Schweidler law of MCT |
| $\tau_{3x}$ | Crossover from non-propagating cage fluctuations to α-process | Crossover from non-stationary to stationary non-phononic modes |
| $\tau_n$ | Eq. (15) and (16) | Time scale of non-propagating cage fluctuations $\tau_n \sim \tau_0^2$ |
| $\tau_p$ | Eq. (15) and (17) | Time scale of propagating cage fluctuations $\tau_p \approx$ const. |
| $\tau_\alpha$ | $\tau_\alpha(\phi)=\tau_{SE}(\phi)-\tau_{SE}(\phi_f)$ | α-relaxation time $\tau_\alpha \sim \tau_0^2$ |
| $\tau_{coll}$ | $\sim 1/F_{cage}$ | Collision time |

TABLE II. Characteristic times; definitions and meaning.

ACKNOWLEDGEMENTS

We thank Gary Bryant, Matthias Fuchs, Jürgen Horbach, Miriam Klopotek, Eric Weeks and Martin Oettel for constructive discussions and comments on earlier versions of this paper.

This work was financially supported by the DFG (Grant No. SCHO 1054/7-1)